\documentstyle[12pt,epsf]{article}

\catcode`\@=11   
   
\def\section{\@startsection {section}{1}{\z@}{-3.0ex plus -1ex minus    
    -.2ex}{0.5ex plus .2ex}{\bf }}

\def\abstracts#1{{\centering{\begin{minipage}{13.0truecm}   
        \footnotesize\baselineskip=12pt\noindent   
        \parindent=0pt #1 \end{minipage}}\par}}   
   
\renewenvironment{thebibliography}[1]   
       {\vspace{-3.5ex} \begin{list}{\arabic{enumi}.}   
        {\usecounter{enumi}\setlength{\parsep}{0pt}   
        \footnotesize\baselineskip=12pt\noindent   
        \setlength{\itemsep}{0pt} \settowidth   
        {\labelwidth}{#1.}   
        \settowidth{\leftmargin}{#1.\rule{3mm}{0mm}}   
        \sloppy}}{\end{list}}   
   
\newcommand{\fcaption}[1]{\vspace{1ex}   
        \refstepcounter{figure}   
        \setbox\@tempboxa = \hbox{\footnotesize {\bf Fig.~\thefigure.} #1}   
        \ifdim \wd\@tempboxa > 14cm   
           {\begin{center}   
        \parbox{14cm}{\footnotesize\baselineskip=12pt {\bf Fig.~\thefigure.} #1}   
            \end{center}}   
        \else   
             {\begin{center}   
             {\footnotesize {\bf Fig.~\thefigure.} #1}   
              \end{center}}   
        \fi} 
   
\def\@citex[#1]#2{\if@filesw\immediate\write\@auxout   
        {\string\citation{#2}}\fi   
\def\@citea{}\@cite{\@for\@citeb:=#2\do   
        {\@citea\def\@citea{,}   
        {\csname b@\@citeb\endcsname}}}{#1}}   
\def\@cite#1#2{$\null^{#1}$}

\begin{document}   
   
\begin{center}   
 {\bf   DIRECT ENERGY TRANSFER IN SYSTEMS OF POLYMERIZED ACCEPTORS}\\    
 \vspace{0.5cm} {\footnotesize    
                 G.OSHANIN,  J.DE CONINCK\\    
 {\it    Centre de Recherche en Mod\'elisation Mol\'eculaire, 
Service de Physique Statistique, Universit\'e de
Mons-Hainaut, 20, Place du Parc, 7000 Mons, Belgium}\\
 {\tt   gleb@gibbs.umh.ac.be}\\    
 \vspace{0.2cm}  
 A.BLUMEN\\    
 {\it   Theoretische Polymerphysik,
 Universit\"at  Freiburg,
 Rheinstrasse 12,
79104 Freiburg, Germany}\\    
\vspace{0.2cm}
        S.F.BURLATSKY\\
 {\it  Department of Chemistry 351700, University of Washington, 
Seattle, WA 98195-1700, USA}\\
 \vspace{0.2cm}
         M.MOREAU\\
 {\it Laboratoire de Physique Th\'eorique des Liquides, 
Universit\'e Pierre et
Marie Curie,
  4, place Jussieu, 75252 Paris, France}\\}
\end{center}   
   
\vspace{0.3cm}    
\abstracts{We study  the 
direct incoherent energy transfer from an
immobile excited donor molecule to acceptor molecules, which are
all attached to polymer chains, randomly arranged in a viscous
solvent.  The decay forms are found explicitly, in terms of an
optimal-fluctuation method, for arbitrary confomations of polymers.}  

\vspace{0.5cm} 
   
Long-range transfer of excitation energy from excited donor to acceptor
molecules, i.e. the direct energy transfer (DET), has attracted a great deal
of interest since the pioneering work by F\"orster\cite{forster} on the subject.
A considerable effort has been invested to the analysis of  decay patterns 
in liquid or solid systems with random or regular distributions
of  molecules  and with different types
of transfer channels\cite{alex}. 

Motivated by an extensive technological, experimental
and theoretical interest in the physics and chemistry of heterogeneous systems, 
more recent research focused on the question how the internal morphology of
the material may influence the relaxation laws, or inversely, on the question
whether DET measurements may serve 
for probing both the local and the global (large scale) structural details
of heterogeneous media\cite{klafter}. In this regard, considerable progress
 has been achieved in
the understanding of the relaxation dynamics in certain types of 
heterogeneous systems; the decay forms were found out 
for porous solids and micellar solutions, as well as for systems with
fractal distributions of donor and acceptor molecules\cite{klafter}.

\begin{figure} \begin{center}   
  \fbox{\epsfysize=4cm\epsfbox{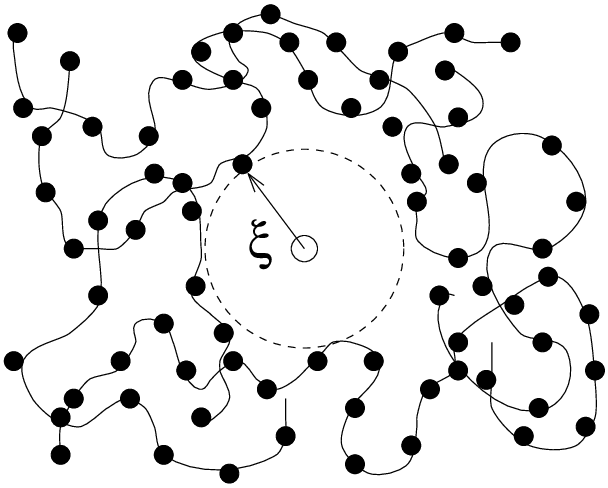}}   
   \fcaption{Direct energy transfer in polymer systems. An open circle
denotes an excited donor molecule, closed circles denote acceptors. The 
parameter
$\xi$ is the radius of the acceptor-free void.}   
  \label{tr}   
\end{center}\end{figure}

In the present paper we address the problem of the donor decay 
in model systems, in which acceptor molecules are all located
on immobile, randomly arranged polymer chains (see ~Fig.~\ref{tr}).  
The theoretical understanding of the relaxation laws in such systems
is practically important for the analysis of  experimental
data on the  DET processes in polymer solutions\cite{win}
and, besides, may provide a deep insight on the relaxation dynamics 
in fractal media,
since polymer solutions exhibit fractal properties on  scales less than
 polymers' gyration radii.  
Exact solutions of this 
and  several related models   
 have been
obtained 
in the particular case of Gaussian  chains\cite{blumen,burl,oshanin}.
Here we focus  on the  general case of chains having arbitrary
conformations and derive explicit forms for the decay laws, employing
the optimal-fluctuation method\cite{lifschitz}. We show that actually 
the exponents
characterizing the donor decay depend on the polymer's conformations, and thus 
the DET measurements may be a source of useful information on the 
intrinsic properties of polymer solutions.

We start with the formulation of the problem and notations. Consider a single 
immobile
donor molecule, located at the origin, and an array of immobile acceptor 
molecules 
placed at positions $\{r_{j}\}$, $r_{j}$ being the radius-vector of the $j$-th 
acceptor.
For the moment we do not specify the geometrical restrictions imposed on 
$\{r_{j}\}$. 
The rate of energy transfer to the $j$-th acceptor molecule is denoted by 
$W(r_{j})$. We 
consider here only the case of isotropic multipolar interactions, for which 
\begin{equation}\label{rate}   
W(r)   =  \frac{W}{r^s},  
\end{equation}   
with, e.g.  $s = 6$ for dipolar and  $s = 10$ for quadrupolar interactions.
Disregarding the
possibility of the back transfer to the donor and assuming that all
 acceptors act independently,
one has that the decay of a donor for a given realization $\{r_{j}\}$
 of the spatial distribution of
acceptors is described by the function
\begin{equation}\label{decay1}   
F(t; \{r_{j}\})   =  exp( - t  \sum_{j}  W(r_{j}) ). 
\end{equation}  
The property which is measured in experiments of
direct energy transfer is not, however, the realization-dependent
 function in (\ref{decay1}),
but rather the global decay function 
\begin{equation}\label{decay2}   
\Phi(t)   =  < exp( - t  \sum_{j}  W(r_{j}) ) >, 
\end{equation}
where the brackets denote the average over all configurations of $\{r_{j}\}$.

Let us adapt now the optimal-fluctuation method to the system under study.
Consider some given realization
$\{r_{j}\}$ and, as depicted in  ~Fig.~\ref{tr}, 
denote as $\xi$ the distance between the donor 
and the nearest acceptor.  
Neglecting next the fluctuations in the spatial distribution of acceptors 
outside
the acceptor-free void (~Fig.~\ref{tr}), we have
\begin{equation}\label{decay3}   
F(t; \{r_{j}\})   \approx  exp( - n_{A} t \int^{\infty}_{\xi} r^{d - 1} dr  
W(r)), 
\end{equation}
where $n_{A}$ denotes the average acceptors' density.       
The radius $\xi$ is a random variable,  and knowing
its distribution function $P(\xi)$ we may estimate the  
function in (\ref{decay2}) as
\begin{equation}\label{opt}   
\Phi(t)   \approx \int d\xi P(\xi)   exp( - n_{A} t \int^{\infty}_{\xi} r^{d - 
1} dr  W(r)). 
\end{equation}

The form of the distribution function $P(\xi)$ depends on the  details of the
acceptors' placement. Particularly, in the situation when all acceptors are 
placed 
at random and independently of each other, 
$P(\xi) \approx exp( - n_{A} \xi^{d})$.
Substituting such a form of $P(\xi)$ into (\ref{opt}) and evaluating the 
integral using
the steepest-descent method, we arrive at the conventional F\"orster-type
 decay law\cite{forster,alex},
$\Phi(t)   \approx  exp( - n_{A} (W t)^{d/s})$.

Next, we briefly discuss how the form of $P(\xi)$ can be derived for
 the situation depicted in ~Fig.~\ref{tr}, when all acceptors are attached to 
polymers
($N$ acceptors per polymer, the length of each polymer is $L$
 and the mean concentration of polymer chains is $n_{p}$, $n_{A} =
n_{p} N$).
A more complete  analysis of this problem can be found in\cite{osha,oshb}. 
In this case, the probability of having an acceptor-free (or, in other words, 
polymer-free)
void of radius $\xi$, can be represented as a product
\begin{equation}\label{dist}   
P(\xi)   =   P_{end}(\xi)  P_{surf}(\xi),
\end{equation}
where the first multiplier defines the probability that a given void
of radius $\xi$ contains no polymer ends, while the second one
defines the conditional probability that no polymer chain,
whose end is placed outside the void, crosses the surface of the void.
Since the polymers' ends are uniformly distributed in space, 
 $P_{end}(\xi) \approx  exp( - n_{p} \xi^{d})$. 
The second multiplier
 can be also found explicitly if one invokes the  analogy
between the conformational statistics of polymers and the statistics of
random walks.  It appears then that $P_{surf}(\xi)$ equals the 
probability that an immobile particle of radius $\xi$ survives until time
$N$ in the presence of a concentration $n_{p}$ of traps performing
random motion, i.e. equals the survival probability in the so-called
"target" problem\cite{ablumen,sfb}.  This gives\cite{osha,oshb}
\begin{equation}\label{distr}   
P(\xi)   \approx   exp( - n_{p} \xi^{d} - n_{p} l^{d_{f}} N \xi^{d - d_{f}}),
\end{equation}
where $l$ denotes the mean distance between the acceptors occupying 
a given polymer chain ($l = L/N$) and $d_{f}$ is the fractal dimension
of polymers, which relates the polymers' radius of gyration, $R_{g}$,
to its contour length $L$, $R_{g} \sim L^{1/d_{f}}$.  For Gaussian
coils $d_{f} = 2$, while for the swollen coils (self-avoiding walks)
$d_{f}$ is inverse to the Flory exponent, $d_{f} = (d + 2)/3$.

Now, inserting (\ref{distr}) into (\ref{opt}) and taking advantage of the
steepest-descent method, we have that in systems with
polymerized acceptors the decay of $\Phi(t)$ displays
two different regimes. At  intermediate times, such that $t < \tau_{g}/N$, 
$\tau_{g} = R_{g}^{s}/W$, 
 the donor decay is described by
\begin{equation}\label{inter}   
\Phi(t)   \approx  exp( - n_{A} l^{d_{f} (1 - z)} ( W  t )^{z}), \;  
\mbox{with} \; z = \frac {d - d_{f}}{s - d_{f}},
\end{equation}
which thus depends explicitly on the polymer conformations.    
For 3D solutions of  Gaussian polymers we get from (\ref{inter}) 
$z = 1/(s - 2)$, ($z = 1/4$ for $s = 6$),  while for rod-like or
swollen polymers (\ref{inter}) yields  $z = 2/(s - 1)$, ($z = 2/5$ for $s = 6$)
and  $z = 4/(3 s - 5)$, ($z = 4/13$ for $s = 6$) respectively. 
At 
large times, $t \gg \tau_{g}/N$, 
the global decay function 
attains the F\"orster-type time dependence
\begin{equation}\label{large}   
\Phi(t)   \approx  exp( - n_{p} (W N t)^{d/s}),
\end{equation}
which is universal, i.e. independent of the conformational properties of
polymer chains.  A dependence such as in (\ref{large}) can be expected on 
physical grounds,
since the long-time behavior of $\Phi(t)$ mirrors the large scale properties
of polymer solutions, when, due to the fact that all polymers are of a finite 
extent, each
chain can be viewed as being an "effective" point acceptor of strength $W N$.
We note, however, that the decay in (\ref{large}) will be accessible 
experimentally
only for very diluted polymer solutions, when $n_{p} R_{g}^{d} \ll 1$. In the
opposite case,  the bulk of excited donors will be deactivated at the 
intermediate time regime,
described by (\ref{inter}).

\section*{Acknowledgements}   
A.B. acknowledges financial support by the Deutsche Forschungsgemeinschaft (SFB 
60)
and by the Fonds der Chemischen Industrie; G.O. acknowledges support from
the FNRS, Belgium;  J.D.C. and M.M. are partially supported by the COST Project 
D5/0003/95.  
The research of S.F.B. is supported by the ONR Grant N 00014-94-1-0647
and by the PAST Programme of the French Ministry of Education.

\end{document}